\documentclass[10pt,preprint]{aastex}  

\makeatletter
\@ifundefined{lesssim}{}{}
\@ifundefined{gtrsim}{\def\gtrsim{\mathrel{\mathpalette\vereq>}}}{}
\def\vereq#1#2{\lower3pt\vbox{\baselineskip1.5pt \lineskip1.5pt
\ialign{$\m@th#1\hfill##\hfil$\crcr#2\crcr\sim\crcr}}}
\makeatother

\shorttitle{Chaotic motion of Charged Particles}
\shortauthors{Takahashi \& Koyama}

\begin{document}


\title{ Chaotic motion of Charged Particles  
        in an Electromagnetic Field Surrounding a Rotating Black Hole } 


%

\author{Masaaki Takahashi}%
\affil{%
 Department of Physics and Astronomy, Aichi University of Education,  
 Kariya, Aichi 448-8542, Japan 
}%
\email{takahasi@phyas.aichi-edu.ac.jp} 

\and


\author{Hiroko Koyama}%
\affil{%
 Department of Physics, Nagoya University, Nagoya 464-8602, Japan  
}%
\affil{%
 Department of Physics, Waseda University, Shinjuku-ku, Tokyo 169-8555, Japan
}%




\begin{abstract}
 The observational data from some black hole candidates suggest the
 importance of electromagnetic fields in the vicinity of a black hole. 
 Highly magnetized disk accretion may play an importance rule, and
 large scale magnetic field may be formed above the disk surface. 
 Then, we expect that the nature of the black hole spacetime would be
 reveiled by magnetic phenomena near the black hole. 
 We will start to investigate the motion of a charged particle which
 depends on the initial parameter setting in the black hole dipole
 magnetic field. 
 Specially, we study the spin effects of a rotating black hole on the
 motion of the charged particle trapped in magnetic field lines.   
 We make detailed analysis for the particle's trajectories by using the
 Poincar\'{e} map method, and show the chaotic properties that depend 
 on the black hole spin. 
 We find that the dragging effects of the spacetime by a rotating black
 hole weaken the chaotic properties and generate regular trajectories
 for some sets of initial parameters, while the chaotic properties
 dominate on the trajectories for slowly rotating black hole cases.  
 The dragging effects can generate the fourth adiabatic invariant on the
 particle motion approximately.   
\end{abstract}



\keywords{%
 magnetic field --- black hole physics --- relativity, chaos 
}%


\section{Introduction}

 In the central region of AGNs, compact X-ray sources and GRBs, many
 black hole candidates are reported.  The observational data from their 
 active regions would include the informations about the gravitational
 and electromagnetic fields.  However, there are no direct observational
 evidences for the existence of a black hole in the center of the
 compact active regions.  So, it is very important how to find the
 informations on the curved spacetime from observational data, and how 
 to interpret their nature.   
 Although the inner part of a gas disk around a supermassive black hole
 emits non-thermal radiation, this may be explained by synchrotron
 emission from ultra-relativistic electrons in strong magnetic field
 regions around a black hole.  In the central black hole system,
 however, the emission process is very little understood, although the
 consensus is in favor of emission due to plasma process.  The emission
 from such particles would give us the information about the black hole
 spacetime and/or the electromagnetic field.

 In a realistic model of a black hole--disk system, we should consider
 interactions between a plasma particle and a magnetic field in the
 curved spacetime.  The plausible electromagnetic field around a black
 hole--disk system is still unknown, although some authors have
 discussed this problem in the frame work of magnetohydrodynamics
 \citep[e.g.,][]{BZ77,Camenzind87,NTT91,TT01}.  In general, the task for
 solving the structure of magnetic fields in a black hole system is very
 hard.  Then, to make a comprehensive analysis, we will give a 
 {\it vacuum} external magnetic field around a rotating black hole, and
 investigate in detail the motion of a charged  particle confined in the
 magnetic field.  The stationary and axisymmetric vacuum electromagnetic
 fields around a black hole are discussed by
 \cite{Pett75,CV75,Li00,Ghosh00,TT01}.   
 We apply the black hole dipole magnetic field in Kerr geometry, which
 is a solution of the vacuum Maxwell's equations.  Although the dipole
 magnetic field could correspond to the intrinsic field of a compact
 object like a neutron star, in the case of black holes it has to
 results from current rings exterior to the event horizon but they can
 be very close to it \citep{Prasanna78}.      
 As known in the Earth's magnetosphere, the dipole magnetic field can
 trap charged particles within its magnetic bottle (i.e., the Van
 Allen belts).  The particle gyrates around the magnetic field line, 
 and drifts in the toroidal direction \citep[see, e.g.,][]{GP04}.
 Furthermore, in the poloidal plane, the particle oscillates along the
 magnetic field line.  In the black hole case, we can also expect
 similar situations.

 In a black hole magnetosphere, 
 by solving equations of motion of a charged particle moving in the
 electromagnetic field numerically \citep[see][]{Prasanna78}, we can 
 see the gyration around the magnetic field line, the drift in the
 toroidal direction and the multi-periodic motion in the poloidal
 plane.  Such a orbit is very complicated in both poloidal and toroidal
 directions.   
 Although a stationary and axisymmetric magnetosphere is assumed, there
 are two conserved quantities; that is, the energy and angular momentum 
 of the charged particle.  In addition to these, the rest mass of the
 particle is the third invariable quantity.  However, the fourth
 invariable quantity related to the azimuthal motion of the particle 
 does not exist when the electromagnetic field is considered around a
 black hole.  Thus, the motion of a test charge in a black hole
 magnetosphere is not an integrable system.  Then, the chaotic motion 
 is expected \citep{NI93}.  
 To examine the trajectory, we plot the Poincar\'{e} map in the
 two-dimensional $r$-$p^r$ plane, which shows intersections of the
 particle's trajectories with the surface of section in phase space
 \citep{LL92}.    
 If the plotted points form a closed curve, the motion is regular (not 
 chaotic).  This is because a regular trajectory moves on a torus in the
 phase space and the curve is a cross section of the torus.  On the
 other hand, if the plotted points are distributed randomly, the motion
 is irregular (chaotic).  From the distribution of the points in the 
 Poincar\'{e} map, we can judge whether or not the motion is chaotic. 
 Then, we know chaotic behavior on the trajectory of a charged particle 
 in a black hole magnetosphere.  Furthermore, we find the black hole's 
 spin effects on that chaotic motion.

 This study will be related to the source of radiation near a black hole.  
 We will start this work as a basic step before considering the plasma 
 as a magnetized fluid.  Although we find chaotic and/or regular motion
 in a black hole magnetosphere, the problem is how to find the phenomena
 in the observed spectrum.  At this stage, however, we can not show the
 relation with the observational data in this work.  
 The main purpose of this study is to understand the black hole spin
 effects (the dragging effects by a rotating black hole) on a test
 charge, which will be considered as the source plasma of high energy
 radiation in future.

 In \S~\ref{sec:basic-eqs}, we review the basic equations for the
 motion of a charged particle in a black hole magnetosphere,
 using the Hamilton-Jacobi formalism.   
 To understand the nature of orbits of a charged particle in the black
 hole magnetosphere, we also consider in \S~\ref{sec:orbit-mag} the
 effective potential including the four-vector potential of the magnetic
 field. We present the motions of a test charge in the black hole dipole
 magnetic field for various values of black hole's spin parameters.   
 In \S~\ref{sec:chaos} we make detailed analysis for its trajectories by 
 using the Poincar\'{e} map method, and show the chaotic properties that
 depend significantly on the black hole spin obviously. 
 We see simultaneous presence of regular trajectories and regions of
 stochastic trajectories on the Poincar\'{e} map.  Furthermore, we find
 that due to the dragging effects of a rapidly rotating black hole, the
 fourth constant of motion {\it approximately}\/ turns out to exit.   
 Then, the motion of a test charge in a dipole magnetic field around a
 rotating black hole can behave like a integrable system.

\section{ Equations for a Charged particle around a Black Hole } 
\label{sec:basic-eqs} 

 We study the motion of a relativistic charged particle with the rest
 mass $m$ and charge $q$ in Kerr geometry.  The background metric is
 written by the Boyer-Lindquist coordinates ($t, r, \theta, \phi$) with
 $c=G=1$, where the signature of the metric is $-2$.  The non-zero
 components of the contravariant metric $g^{\mu\nu}$ are given by   
\begin{equation}  
   g^{tt}=\frac{A}{\Delta\Sigma}\ , ~~~  
   g^{t\phi}=\frac{2Mar}{\Delta\Sigma}\ , ~~~ 
   g^{\phi\phi}= - \frac{1-2Mr/\Sigma}{\Delta\sin^2\theta} \ , ~~~ 
   g^{rr}= - \frac{\Delta}{\Sigma} \, ~~~  
   g^{\theta\theta}= - \frac{1}{\Sigma} \ . 
\end{equation} 
 where $\Sigma \equiv r^2+a^2\cos^2\theta$, $\Delta \equiv r^2-2Mr+a^2$,   
 $A \equiv (r^2+a^2)^2 - \Delta a^2 \sin^2\theta$.  
 The external electromagnetic field is assumed to be stationary and 
 axisymmetric. We assume that the electromagnetic fields would not 
 disturb the background geometry of the spacetime. In fact the 
 electromagnetic fields seen in astrophysical black hole candidates 
 would be quite small compared with the gravitational field associated 
 with the black hole.

 The Hamiltonian of the system can be defined as \cite[e.g.,][]{MTW73}  
\begin{equation} 
   H \equiv \frac{1}{2}g^{\mu\nu}(\pi_\mu - q A_\mu)(\pi_\nu - q A_\nu)  
                                                 \label{eq:hamiltonian}
\end{equation} 
 where $\pi_\mu$ is the canonical momentum and $A_\mu=A_\mu(r,\theta)$ 
 is the four-vector potential of the electromagnetic field.  
 The four-momentum of a charged particle is   
\begin{equation} 
   p^\mu \equiv \frac{dx^\mu}{d\lambda} 
              = g^{\mu\nu}(\pi_\nu - q A_\nu)  \ ,  
\end{equation} 
 where $\lambda=\tau/m$ is an affine parameter and $\tau$ is the
 proper time.  The Lorentz-force equation is equivalent to Hamilton's
 equations written in terms of $x^\mu$ and $\pi_\mu$: 
\begin{eqnarray} 
  \frac{dx^\mu}{d\lambda} &=& \frac{\partial H}{\partial \pi_\mu}   \ , 
                                         \label{eq:heq-1}  \\ 
  \frac{d\pi_\mu}{d\lambda} &=& - \frac{\partial H}{\partial x^\mu} \ .  
                                         \label{eq:heq-2}  
\end{eqnarray}

 In a stationary and axisymmetric black hole magnetosphere, the magnetic
 field is specified by a scalar function $\Psi$ of position called the
 stream function; that is, the poloidal magnetic field lines are
 contours of constant $\Psi$. 
 (The function $\Psi$ is proportional to the toroidal component of the
 vector potential, $A_\phi$.) 
 In this paper we set up a vacuum magnetosphere, so that no total
 poloidal current exists, and then the toroidal component of the
 magnetic field is zero.

 From the stationary and axial symmetry of both electromagnetic field
 and spacetime geometry, $ E \equiv \pi_t = p_t + q A_t$ and 
 $L \equiv -\pi_\phi = -(p_\phi + q A_\phi)$ are constants of motion
 corresponding to the integrable coordinates $t$ and $\phi$ in
 eq.~(\ref{eq:hamiltonian}).  The third constant of motion is the 
 particle's rest mass $m = (-g^{\mu\nu}p_\mu p_\nu)^{1/2}$.  
 In general, {\it four}\/ constants of motion are needed to determine
 uniquely the orbit of a particle through four-dimensional spacetime.
 However, a test charge motion in a magnetic field around a black hole
 posses only {\it three}\/ obvious constants.  So we can expect chaotic
 behavior for its motions \citep{LL92,KV92}, because such a system is
 non-integrable.   
 Note that, without the magnetic field, the fourth constant of motion 
 exists.  This constant is known as Carter's constant of the motion
\begin{equation}
   {\cal Q} = p_\theta^2 +\cos^2\theta \left[
             a^2(m^2 - E^2) + \frac{L^2}{\sin^2\theta} \right] \ ,
   \label{eq:Carter}
\end{equation}
 which arises as a separation-of-variables constant in the Hamilton-Jacobi
 derivation of equation of motion~\citep[see][]{MTW73}. That is, the
 motion of a test particle in Kerr spacetime without magnetic field is
 an integrable system.

 The stationary and axisymmetric electromagnetic fields around a
 rotating black hole in source-free regions are derived by
 \cite{Pett75}.  Here, we consider the solution that denotes dipole  
 magnetic fields at distant regions, and call this solution ``black hole
 dipole electromagnetic field''.  In this black hole magnetosphere, we
 expect that the dipole magnetic field configuration can trap a charged
 particle within their magnetic bottle, even if it extends close to the
 black hole. 
 The four-vector potential of the black hole dipole has only two nonzero 
 components, $A_t$ and $A_\phi$; the dipole magnetic field and an
 induced quadrupole electric field in Kerr geometry are given by   
\begin{eqnarray} 
         A_t &=& \frac{-3a\mu}{2\gamma^2\Sigma}\left\{ 
        \left[ r(r-M) + (a^2-Mr)\cos^2\theta \right] \frac{1}{2\gamma} 
        \ln\left(\frac{r-r_{-}}{r-r_{+}}\right) \right.  \nonumber \\  
        &~& ~~~~~~~~-(r-M\cos^2\theta) \bigg\}  ~\ , 
         \label{eq:dipole-kerr_At}  \\ 
        A_\phi &=& \frac{-3\mu\sin^2\theta}{4\gamma^2\Sigma} \bigg\{ 
        (r-M)a^2\cos^2\theta + r(r^2+Mr+2a^2)            \nonumber \\  
        &~& ~~~~~~~ -\left[ r(r^3-2Ma^2+a^2r) + \Delta a^2\cos^2\theta  
         \right] \frac{1}{2\gamma} 
        \ln\left(\frac{r-r_{-}}{r-r_{+}}\right)  \bigg\} \ , 
         \label{eq:dipole-kerr_Af} 
\end{eqnarray} 
 where $\gamma\equiv (M^2-a^2)^{1/2}$, $r_\pm = M\pm \gamma$ and the
 dipole moment $\mu$ is taken to be anti-parallel to the rotation axis.  
 In the $a\to M$ limit, we obtain 
\begin{eqnarray} 
   A_t    &=& \frac{ - M\mu [ r\sin^2\theta - 2(r-M)\cos^2\theta ]} 
                   { 2 (r^2+M^2\cos^2\theta)(r-M)^2} \ ,
                                                    \label{eq:At_M} \\  
   A_\phi &=& \frac{ - \mu\sin^2\theta  
                   [ - 2 r^3 + M (r-M) (r + M \cos^2\theta) ]} 
                   { 2 (r^2+M^2\cos^2\theta)(r-M)^2 } \ .              
                                                    \label{eq:Af_M} 
\end{eqnarray} 
 In the limit of $a\to M$, $|A_t|$ and $|A_\phi|$ diverge as
 $(r-M)^{-2}$, while in the case of $a<M$, $|A_t|$ and $|A_\phi|$
 diverge logarithmically at the event horizon. 
 Although the source of the magnetic field should be outside the event 
 horizon, the current loops could locate just outside the event
 horizon (arbitrarily close to it) as the source of such dipole magnetic
 fields as considered in this work \citep{Prasanna78}.      
 In spite of this singularity, we will use this configuration for the
 study of the chaotic motion of a test charge around a black hole.  
 The appearance of the singularity does not imply that the dipole field
 is invalid.  It is valid in the regions considered outside the event
 horizon.  In a realistic situation for the black hole magnetosphere as
 an astrophysical model, to avoid the singularity at the event horizon,
 the infinite sum of multi-pole fields is necessary \citep{Li00,TT01}.  
 The plausible magnetic field configuration around a black hole should
 be investigated as a future work.

\section{ Orbits of a charged particle in a magnetosphere } 
\label{sec:orbit-mag} 

 Now, we will discuss charged particle motions off the equatorial
 plane of the black hole dipole field.  The concrete expressions of
 equations (\ref{eq:heq-1}) and (\ref{eq:heq-2}) including
 electromagnetic field terms are very complicated, and are not
 particularly informative.  However, by considering an effective
 potential in the poloidal plane, we can get a general picture of  
 the orbits, although we need to integrate the equations of motion
 numerically to obtain the practical orbits. 
 
 \begin{figure}
  \epsscale{.90}
   \plotone{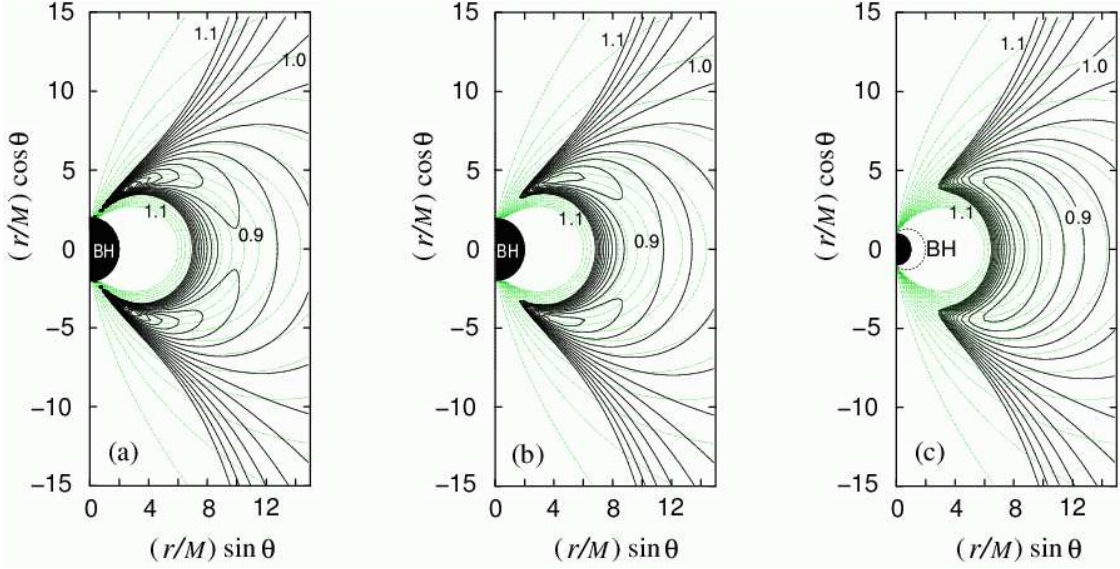}
  \caption{ 
  Effective potential with the black hole dipole magnetic field (thin
  gray curves).   
  (a) $a=0.0$, (b) $a=0.3M$ and (c) $a=M$, where $L/m=-7.0M$ and
  $Q_{d}=70.0M^2$.  In these plots, the maximum level of $V_{\rm eff}$
  is 1.1 and the interval is 0.02.
  }
  \label{fig:poten} 
 \end{figure} 
 
 \begin{figure}
  \epsscale{1.05} \plottwo{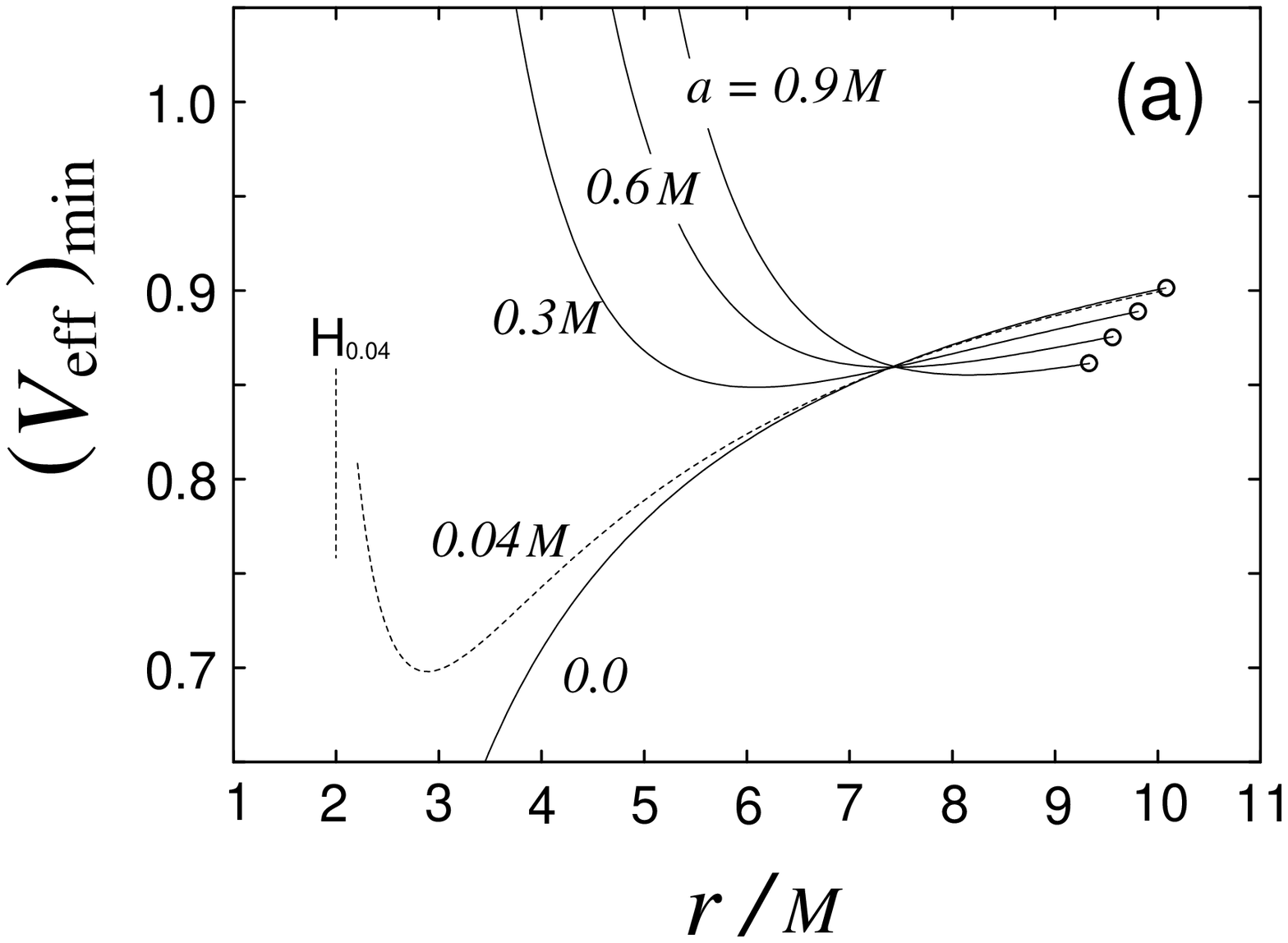}{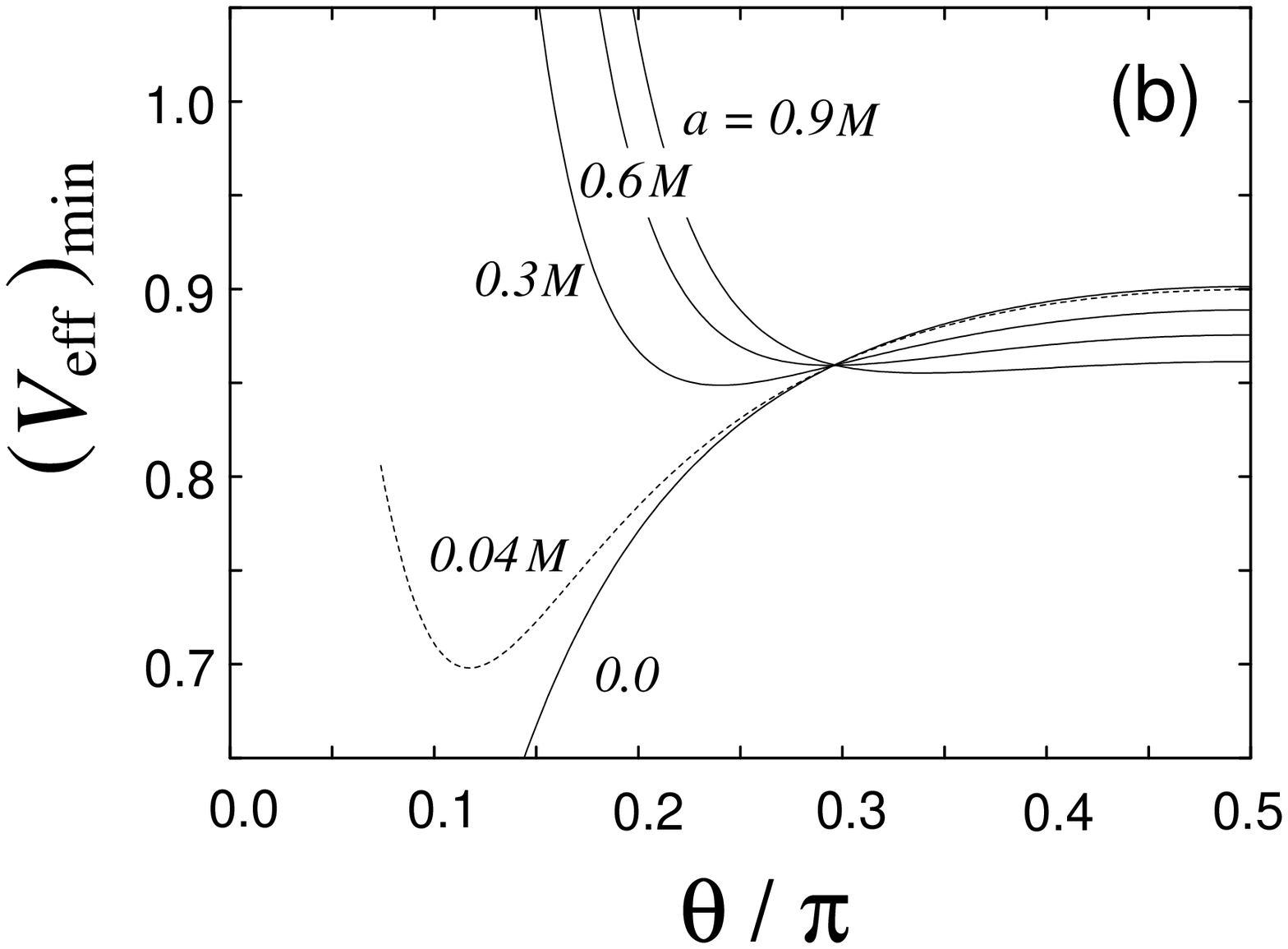}
  \caption{ 
  Potential valley for various spin values. 
  (a) $(V_{\rm eff})_{\rm min}$ vs. $r$ and  
  (b) $(V_{\rm eff})_{\rm min}$ vs. $\theta$ are plotted,   
  where $L/m=-7.0M$ and $Q_d=70.0M^2$. The circles in (a) indicate
  $\theta=\pi/2$, and the vertical line labeled by H$_{\rm 0.04}$
  shows the event horizon for $a=0.04M$. 
  } 
  \label{fig:pot-min} 
 \end{figure} 

 Here, we consider off-equatorial motion for a charged particle in the
 dipole magnetic field.   
 By using the condition $g^{\mu\nu}p_\mu p_\nu=m^2$ with the constants 
 of motion $E$ and $L$, the equation of motion in the poloidal plane can
 be obtained.  Then, without the kinetic terms of the poloidal motion   
 ($p^r=p^\theta=0$), we can define the effective potential \citep{EM90}
 as   
\begin{equation} 
  V_{\rm eff}(r,\theta) \equiv \frac{E_{\rm min}}{m}  
     = \frac{q}{m}A_t + \frac{g^{t\phi}}{g^{tt}} 
            \left(\frac{L}{m}+\frac{q}{m}A_\phi \right) 
      + \frac{1}{g^{tt}}\left[ \frac{1}{\rho_w^2} \left( \frac{L}{m} 
         +\frac{q}{m}A_\phi \right)^2 + g^{tt} \right]^{1/2} \ ,  
   \label{eq:potential} 
\end{equation} 
 where $E_{\rm min}$ is the allowed minimum energy for a particle at
 the injection point.  Although the distribution of the effective
 potential depends on the value of $L/m$ and $q/m$ of the charged
 particle and the electromagnetic field configuration $A_t$ and
 $A_\phi$, to study the chaotic behavior we are interested in the case 
 that a charged particle is trapped in a bound orbit.    
 When the energy of a charged particle is not so large ($V_{\rm eff} <
 E/m < 1$), the particle is bound in the orbit in the region of the 
 potential well.  That is, the particle has turning points, which
 correspond to the inner and outer envelope of the Larmor motion in the
 poloidal plane.

 The detailed motion in the equatorial plane ($p^\theta=0$) in the Kerr
 background with the dipole magnetic field has been discussed by
 \cite{Prasanna80}.  We will also see the projection onto the equatorial 
 plane of the off-equatorial particle's motions.   
 From the $\phi$-component of the equation of motion, we obtain 
\begin{equation} 
    p^\phi = g^{t\phi}(E-qA_t) - g^{\phi\phi}(L+qA_\phi) \ . 
\end{equation} 
 For a non-rotating black hole case, the first term is negligible. 
 When the signatures of parameters $L$ and $q\mu$ are same, the particle 
 does not gyrate because of $p^\phi\neq 0$.  To observe the gyration
 motion of a charged particle, where $p^\phi=0$ is achieved
 periodically, we will choose $L<0$ and $Q_d\equiv (q/m)\mu>0$ in the
 following numerical calculations.  
 To see the off-equatorial motion of a charged particle, the initial
 value of $p^r$ or $p^\theta$ should be specified.  In this paper,  
 we inject the particle into the magnetosphere with the injection angle
 $\psi_{\rm inj}$, which is the angle from the origin of the coordinate
 axes in the poloidal plane and is relate to the ratio of the initial
 $p^r$ and $p^\theta$-values, where we choose $ 0 < \psi_{\rm ini} <
 0.5 \pi$ for the calculations.  The label ``ini'' indicate the
 quantity at the injection point.  
 The equations are solved by using the 4th-order Runge-Kutta-Gill
 method.

 Figure~\ref{fig:poten} shows the hole's spin dependence on the
 effective potential $V_{\rm eff}(r,\theta)$ for the poloidal motion of
 a charged particle in the black hole dipole magnetic field.  We see
 the valley-like structure in the lower-level potential region (named
 ``potential valley''), which is across the northern and southern
 hemispheres nearly along a dipole magnetic field line, wherein the
 particle with lower energy can be trapped inside it.  We also see a
 divide of the potential valley just on the equatorial plane,   
 Figure~\ref{fig:pot-min} shows the $r$- and $\theta$-dependences of the 
 local minimum of the effective potential in the $r$-direction, which is
 obtained from $\partial V_{\rm eff}(r, \theta)/\partial \theta = 0$ and
 corresponds to the bottom along the valley, for various values of the
 spin parameter $a$.  In this plot, we see the local minimum in the 
 middle-latitude region in the $\theta$-direction.

\begin{figure*}[t]
  \epsscale{0.9}
  \plotone{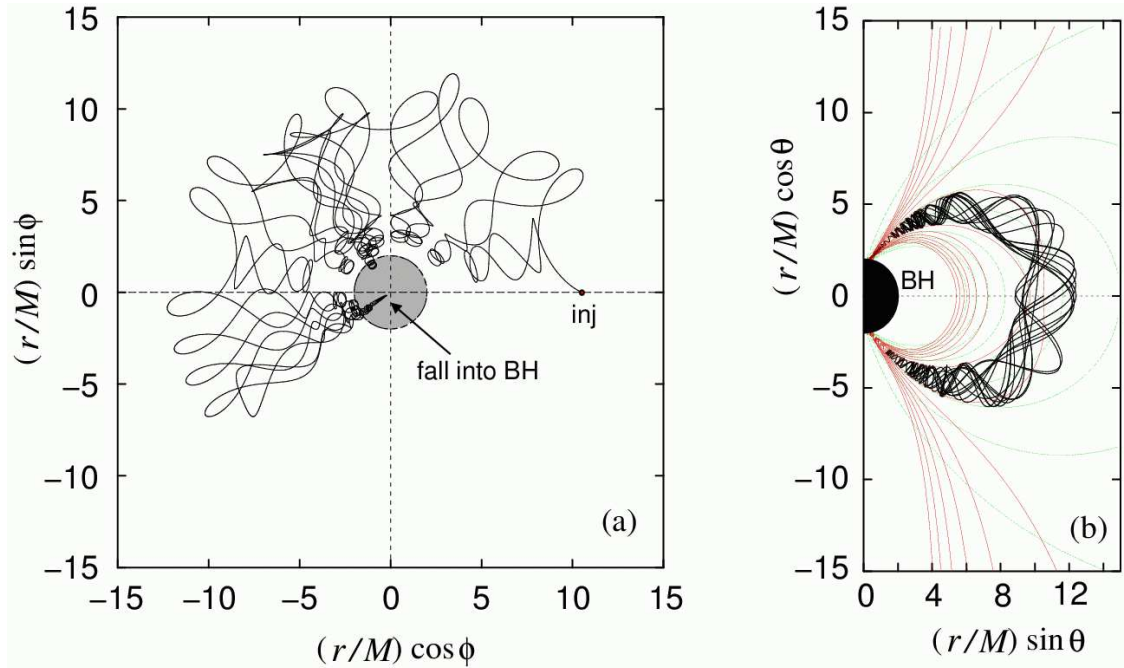}
  \caption{
       The motion of a charged particle in the equatorial plane (LEFT)   
       and in the poloidal plane (RIGHT). The particle finally falls
       into the event horizon from the polar region of the black hole. 
       The parameters of the motion are given as $a=0$, $E/m=0.920$,
       $Q_{d}=70M^2$, $\psi_{\rm ini}=-0.16\pi$, $x_{\rm ini}=10.5M$ 
       and $\theta_{\rm ini}=\pi/2$.   
       In the poloidal plane, the dipole magnetic field lines (thin
       gray curves) and the effective potential (thin curves) are 
       also plotted.      
   }
 \label{fig:non-rot}
\end{figure*}

 In the non-rotating black hole case (see Fig.~\ref{fig:poten}a), the
 value of the local minimum of the effective potential decreases from
 the equator to the event horizon along almost the dipole magnetic field 
 line in both hemispheres, and has the minimum value at the event
 horizon.  
 Thus, the potential valley is opened toward the black hole, while the
 width of the valley becomes narrower toward the event horizon. So,
 although a particle can be trapped within the potential valley for some
 time, the particle will fall into the black hole sooner or later.   
 Figure~\ref{fig:non-rot} is an example of the orbit for a non-rotating
 black hole, where the projection on the equatorial plane (left panel)
 and the poloidal motion (right panel) are shown. 
 The particle gyrates around the magnetic field line, and drifts in the
 toroidal direction.  Furthermore, the particle oscillates in the
 poloidal plane along the magnetic field lines. In this dipole magnetic
 field case, the particle entering the regions of higher magnetic field
 strengths reflects back into the regions of smaller magnetic field
 strength. This is the so-called ``mirror effects''.  Although the
 dipole field traps a particle between two mirrors, the particle falls
 into the black hole in the end, for almost non-rotating black hole
 cases.  In Figure~\ref{fig:non-rot} we show the case of $\psi_{\rm inj}
 = -0.16\pi$, but it is easy to fall into the black hole sooner for the
 larger value of $\psi_{\rm inj}$.

 Next, for a rotating black hole case (see Fig.~\ref{fig:poten}b), the 
 centrifugal barrier by the hole's spin effects arises and the potential
 valley is disconnected from the event horizon.  This is because the 
 centrifugal barrier due to the hole's spin becomes higher with
 increasing the black hole spin.  Thus, the dragging effects of
 spacetime enhance the mirror effects.  Then, the particle can be   
 trapped within the potential valley without falling onto the black
 hole.  Furthermore, we see the ``double-well potential'' along the
 valley.  Basically, the particle oscillates between the northern and
 southern hemispheres, but sometimes a particle with lower energy may be
 trapped in one hemisphere.  Such a double-well potential may be related
 to the  chaotic motion as discussed later.    
 For the extreme rotating case shown in Fig.~\ref{fig:poten}c, the spin
 effects are more effective.  We see an almost single-well potential
 along the valley.  Then, we can expect that the particle would
 oscillate between the northern and southern hemispheres periodically.      
 The trajectory for the particle trapped in the potential valley
 will be discussed in \S \ref{sec:discuss} again.

\begin{figure}[h]
  \epsscale{0.9}
  \plotone{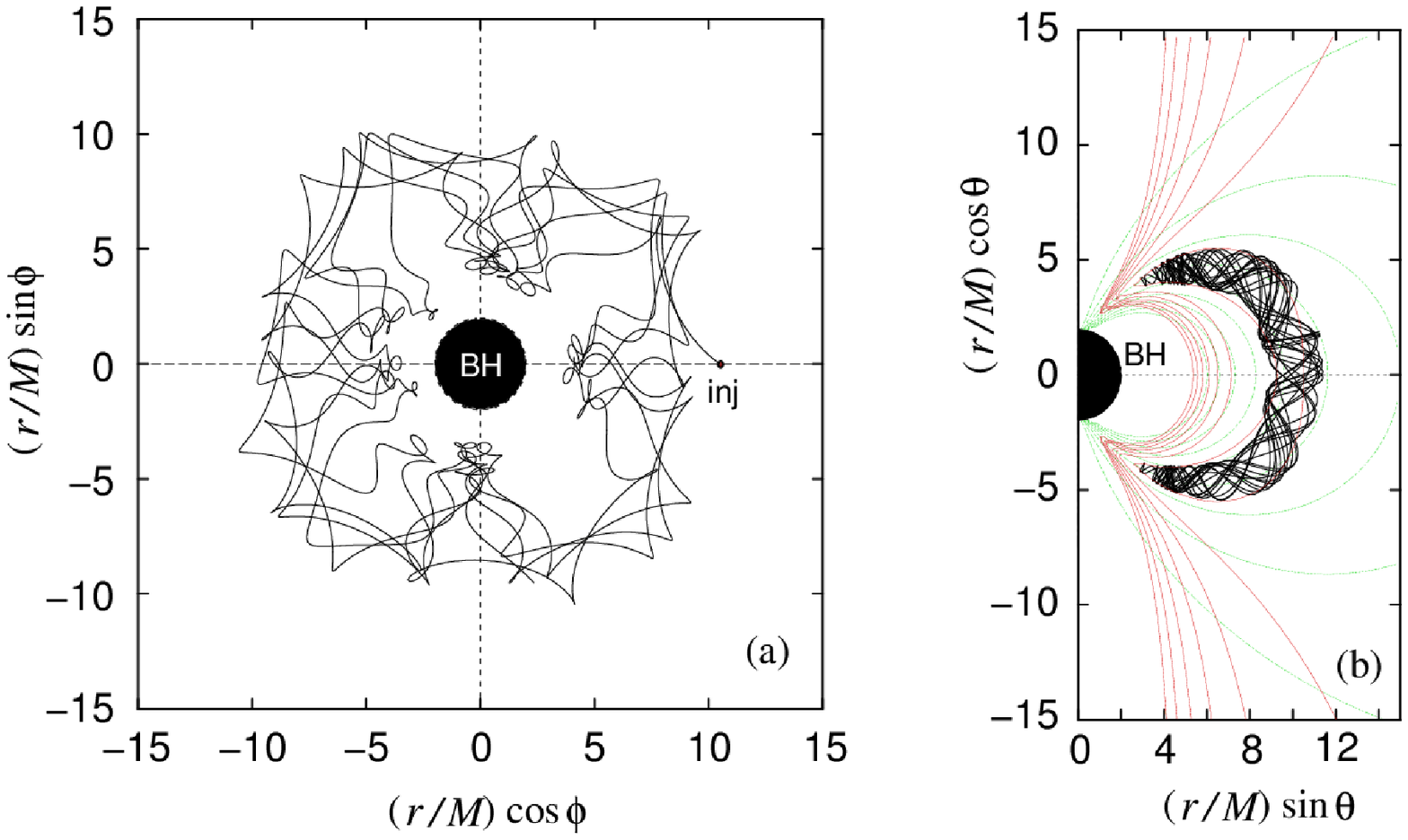}
\caption{
   An Example of the random trajectory of a test charge.  LEFT panel
   shows the projections on the equatorial plane, and RIGHT panel 
   shows the projections on the poloidal plane.  
   The parameters of the motion are set as $a=0.3M$, $E/m=0.900$,
   $\psi_{\rm ini}=-0.20\pi$, $L/m=-7.0M$, $Q_d=70.0M^2$, 
   $r_{\rm ini}=10.5M$ and $\theta_{\rm ini}=\pi/2$.    
}
\label{fig:traj_a3}
\end{figure} 

\begin{figure}
  \epsscale{0.9}
  \plotone{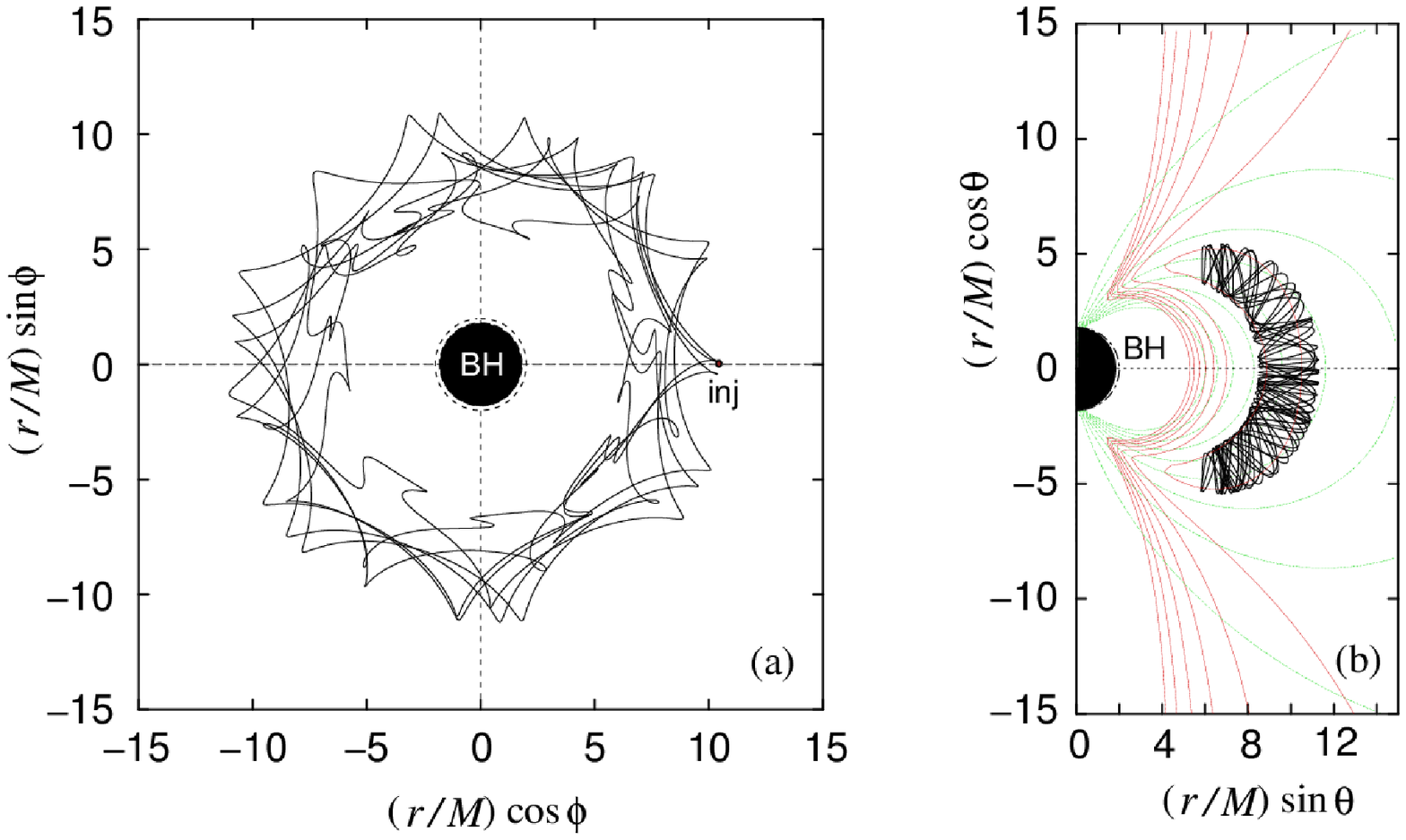}
  \plotone{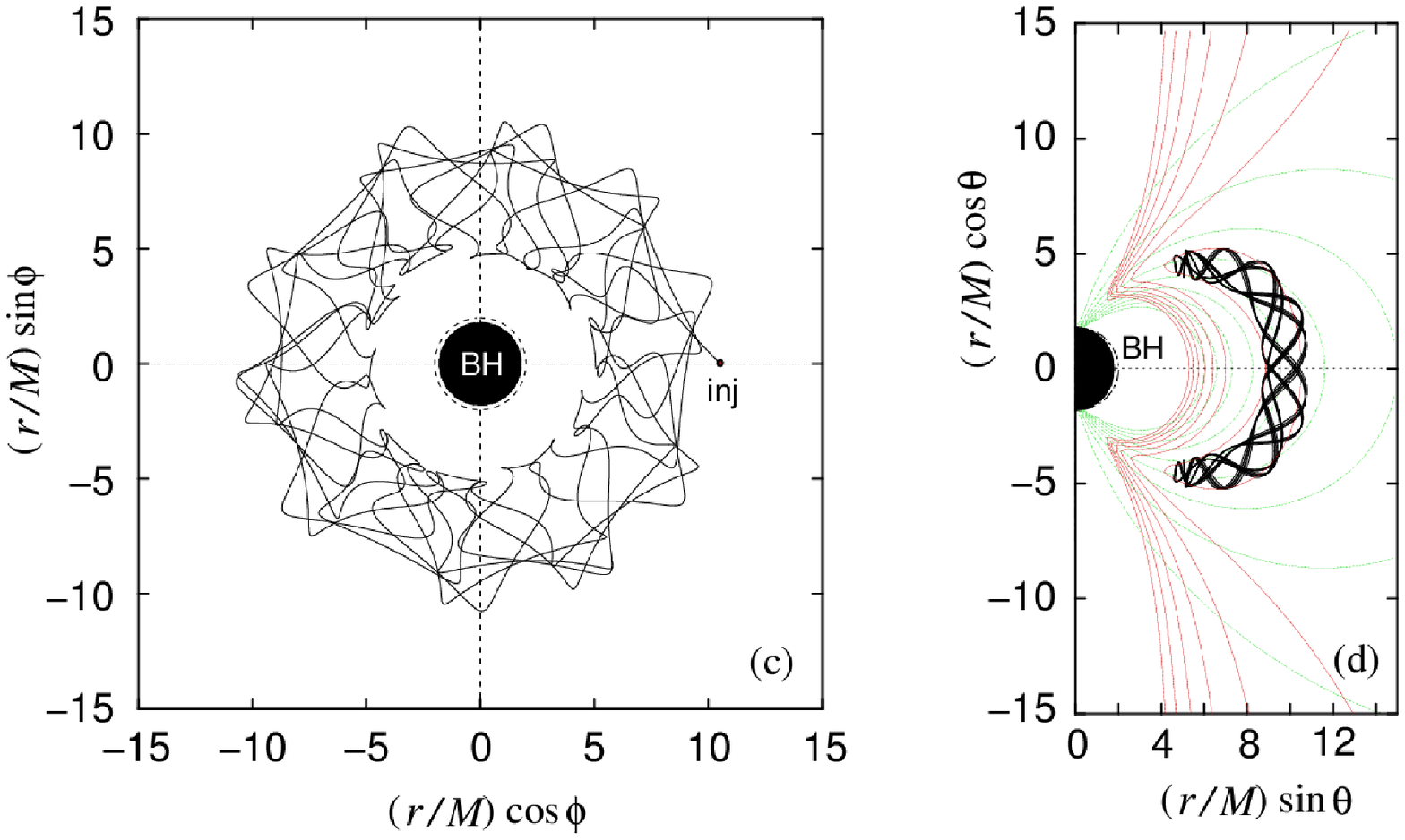}
\caption{
   Examples of the trajectories of a test charge for $a=0.6M$.  
   LEFT column shows the projections on the equatorial plane, and 
   RIGHT column shows the projections on the poloidal plane.  
   The parameters of the motion are set as 
   (a,b) $E/m=0.895$, $\psi_{\rm ini}=-0.10\pi$, 
   (c,d) $E/m=0.890$, $\psi_{\rm ini}=-0.30\pi$, with 
   $L/m=-7.0M$, $Q_d=70.0M^2$, $r_{\rm ini}=10.5M$ and  
   $\theta_{\rm ini}=\pi/2$.   
}
\label{fig:traj_a6}
\end{figure}

\begin{figure}
  \epsscale{0.9}
  \plotone{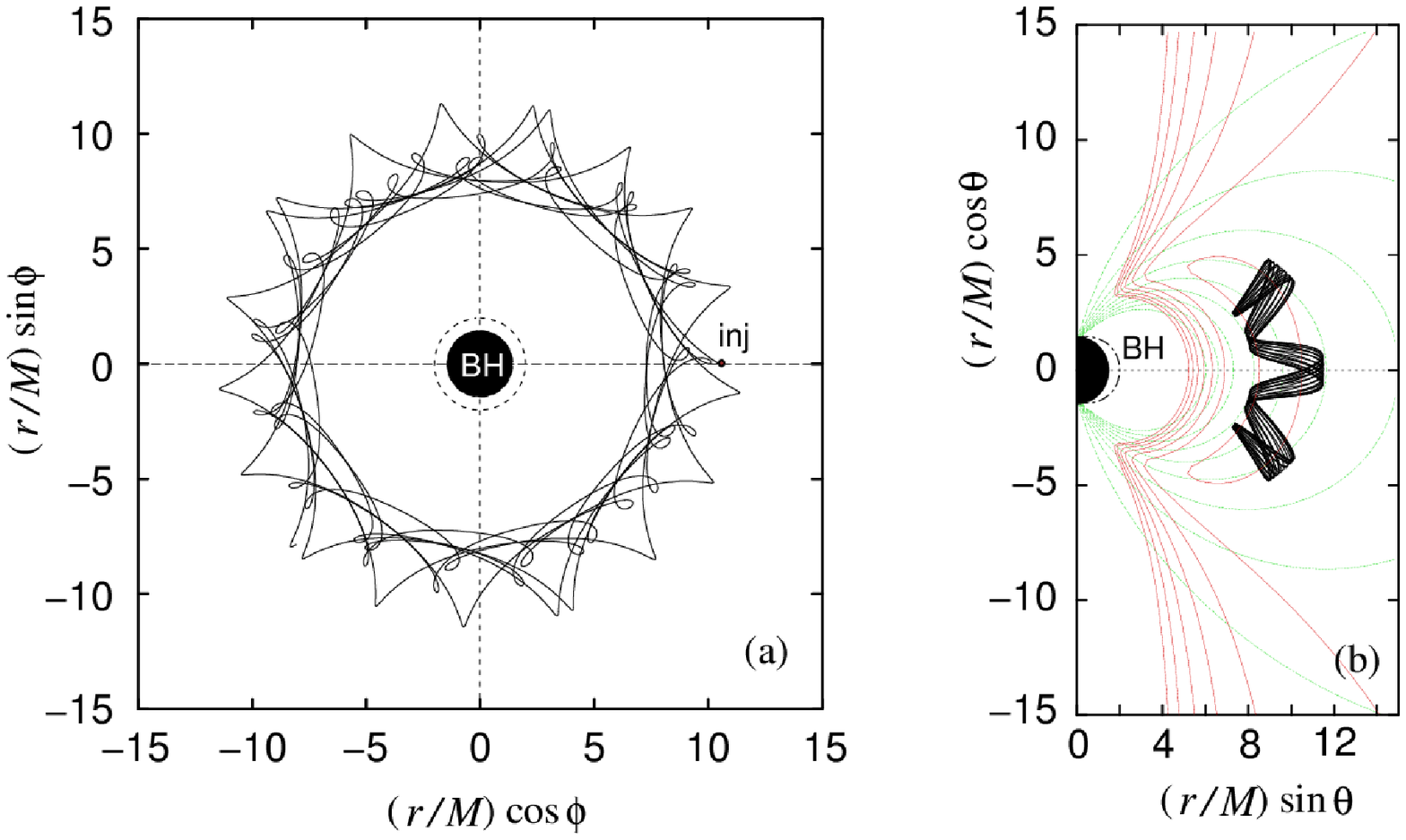}
  \plotone{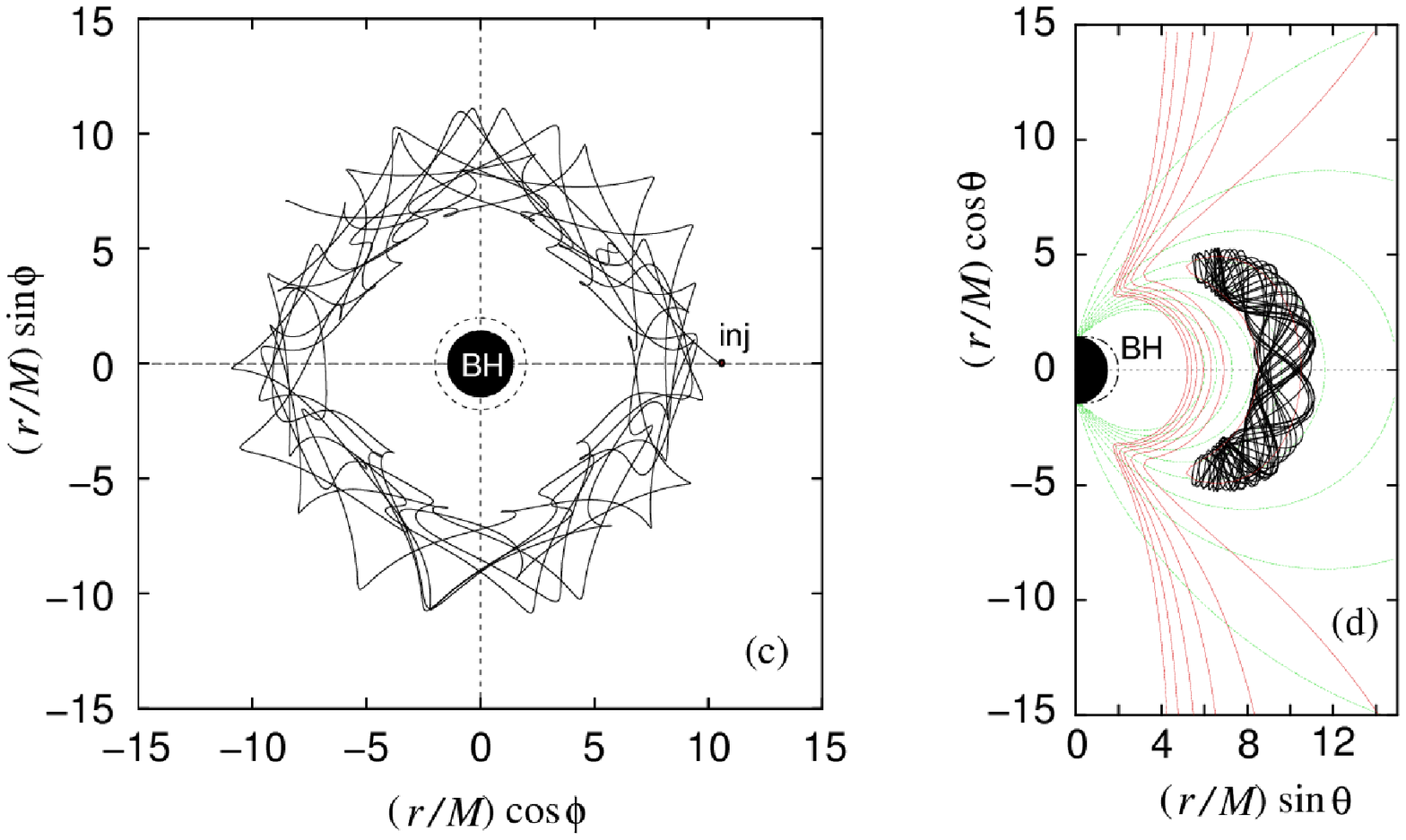}
\caption{
   Examples of the trajectories of a test charge for $a=0.9M$.  
   LEFT column shows the projections on the equatorial plane, and 
   RIGHT column shows the projections on the poloidal plane.  
   The parameters of the motion are set as 
   (a,b) $E/m=0.885$, $\psi_{\rm ini}=-0.10\pi$, 
   (c,d) $E/m=0.885$, $\psi_{\rm ini}=-0.22\pi$, with 
   $L/m=-7.0M$, $Q_d=70.0M^2$, $r_{\rm ini}=10.5M$ and  
   $\theta_{\rm ini}=\pi/2$.   
}
\label{fig:traj_a9}
\end{figure}

 Figures~\ref{fig:traj_a3}--\ref{fig:traj_a9} show the orbits of a
 charged particle around a rotating black hole.  Charged particles
 are trapped within the potential valley (i.e., a magnetic bottle)
 without falling onto a black hole. 
 So, we can carry out long-time calculations for chaos studies.  
 The particle oscillates quasi-periodically between the northern and
 southern hemisphere, although it is a very complicated orbit.  
 In Figures~\ref{fig:traj_a3}, \ref{fig:traj_a6}(TOP)
 and~\ref{fig:traj_a9}(BOTTOM), the particle's orbit seems to be random,  
 while in Figure~\ref{fig:traj_a6}(BOTTOM) and~\ref{fig:traj_a9}(TOP),
 the orbit looks regular; that is, we can see some kind of periodicity
 despite its long-time motion.  For rapidly rotating cases, we can see
 the regular orbits often in comparison with the slowly rotating cases.
 In general, some orbits show a regular trajectory, while a chaotic
 trajectory still remains.  
 Thus, we can find spin effects on the motion of the charged particle.  
 In the following, we analyze the detailed properties of these motions 
 to clarify their dependence on the spin parameter.  Such motions may be
 chaotic.  So, we analyze the properties of the motions using a
 method with which we analyze chaotic behavior.

\section{ Regular and Chaotic orbits }
\label{sec:chaos}

 The particle motions are common in the sense that they are the
 combinations of three types of motions (gyration, bouncing, and
 drifting).  Here we analyze the property of such complicated motions 
 using the Poincar\'{e} map, which shows intersections of a trajectory 
 with the surface of section in phase space.  The motion of a point in
 phase space could be followed over hundreds of thousands of oscillation 
 periods.  The Poincar\'{e} map is an useful tool to classify visually 
 whether the motions are regular (non chaotic) or irregular (chaotic).
 To make the Poincar\'{e} map, we adopt the equatorial plane
 ($\theta=\pi/2$) as a Poincar\'{e} surface and plot the point ($r$,
 $p^r$) when the particle crosses the Poincar\'{e} section with
 ($p^\theta>0$).  
 We obtain that, for a slowly rotating black hole case, a trajectory
 plots a lot of points on the map randomly, while for a rapidly rotating
 case there are some regular trajectories shown by tori.  The different
 tori indicate the different initial injection angle $\psi_{\rm ini}$ on
 the Poincar\'{e} map.  The appearance of regular trajectories suggests
 that the dragging effects of the black hole generate a nearly
 integrable system in spite of the existence of a magnetic field in 
 Kerr spacetime (the details will be discussed in \S~\ref{sec:discuss}).

\begin{figure}
  \epsscale{0.8} \plotone{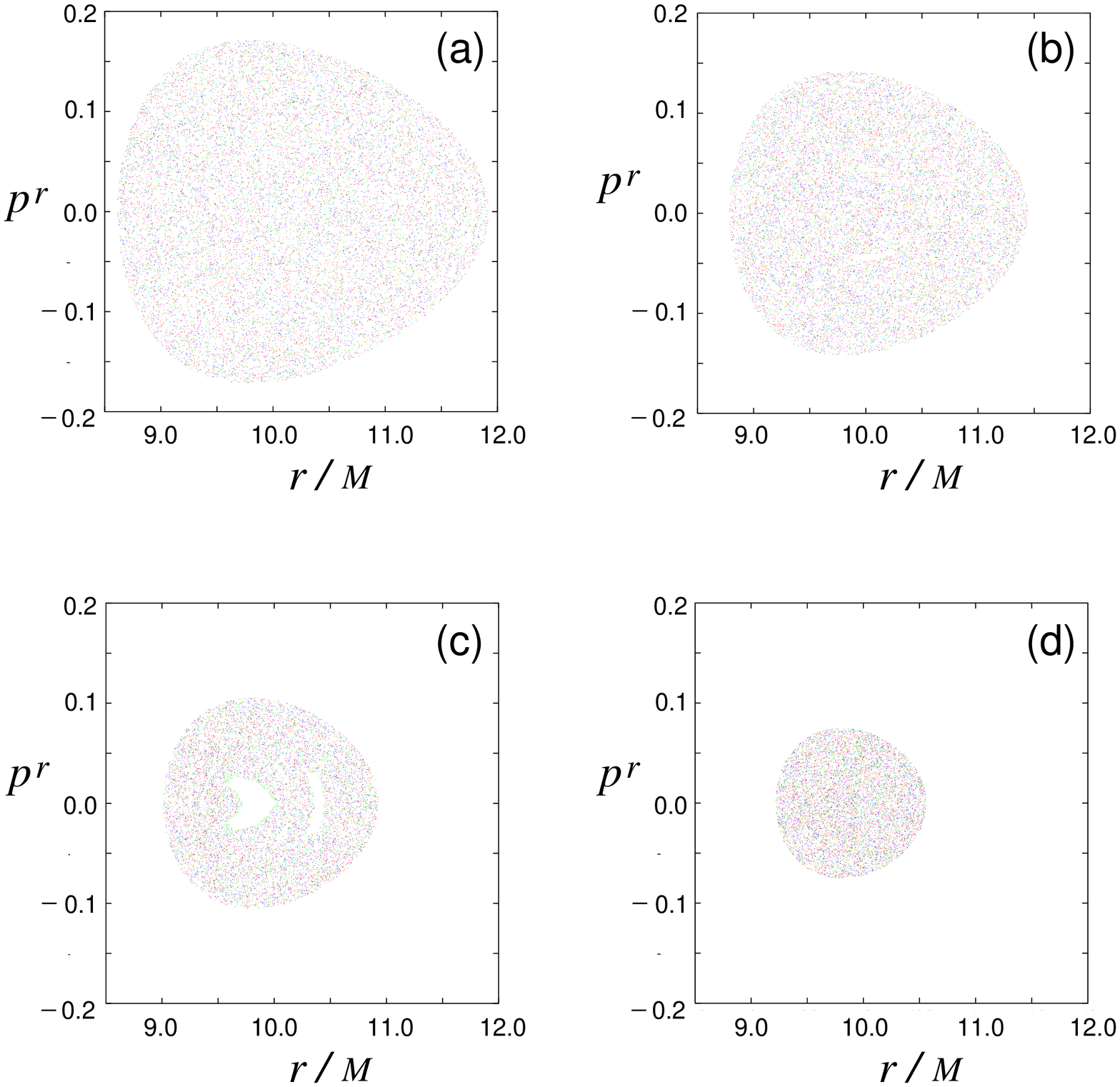}
  \caption{
       The Poincar\'{e} map of a charged particle orbiting in the 
       dipole magnetic field around a black hole of $a=0.3M$. 
       The intersections are shown by small dots on the map.         
       The parameters for the motion are (a) $E/m=0.905$,
       (b) $E/m=0.900$, (c) $E/m=0.895$ and (d) $E/m=0.892$ 
       with $L/m=-7.0M$, $Q_d=70.0M^2$, 
       $r_{\rm ini}=10.5M$ and $\theta_{\rm ini}=\pi/2$. 
   }
 \label{fig:poincare-3} 
\end{figure}

\begin{figure}
  \epsscale{0.8} \plotone{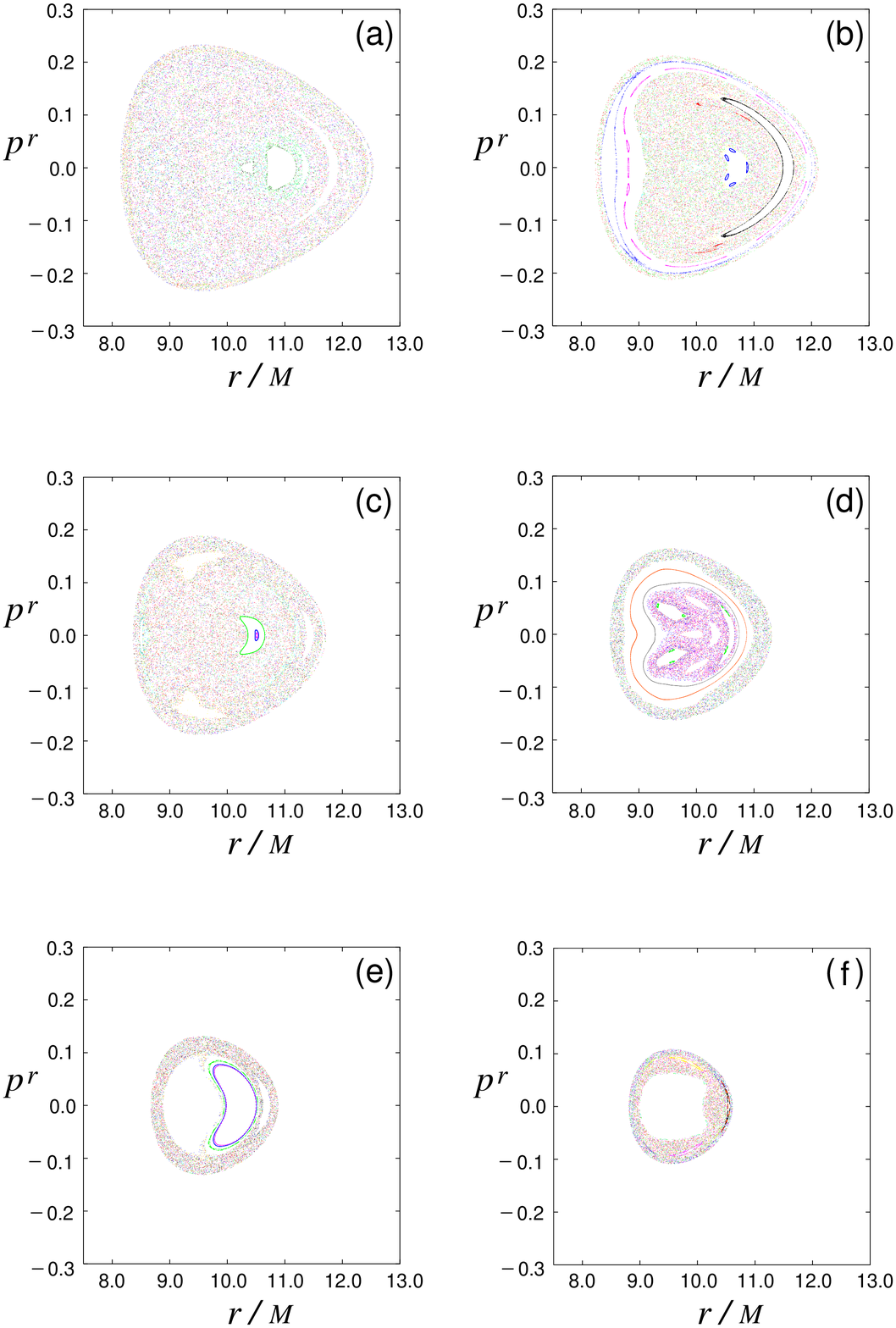}
  \caption{
       The Poincar\'{e} map of a charged particle orbiting in the 
       dipole magnetic field around a black hole of $a=0.6M$. 
       The parameters for the motion are (a) $E/m=0.905$, 
       (b) $E/m=0.900$, (c) $E/m=0.895$, (d) $E/m=0.890$, 
       (e) $E/m=0.885$ and (f) $E/m=0.882$ 
       with $L/m=-7.0M$, $Q_d=70.0M^2$, $r_{\rm ini}=10.5M$ and  
       $\theta_{\rm ini}=\pi/2$.  
   }
 \label{fig:poincare-6} 
\end{figure}

\begin{figure}
  \epsscale{0.8} \plotone{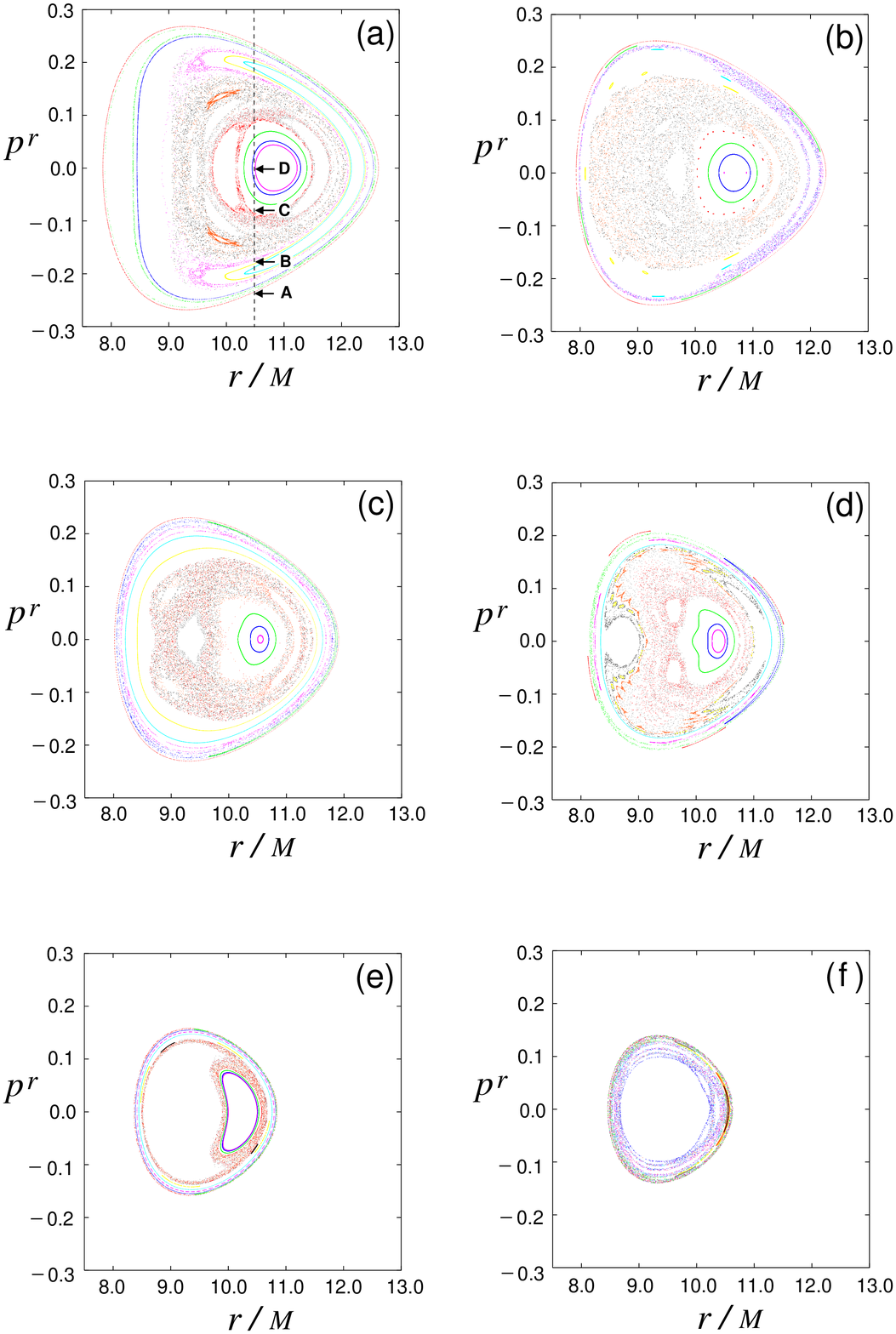}
  \caption{
       The Poincar\'{e} map of a charged particle orbiting in the 
       dipole magnetic field around a black hole of $a=0.9M$. 
       The parameters for the motion are 
       (a) $E/m=0.900$, (b) $E/m=0.895$, (c) $E/m=0.890$, 
       (d) $E/m=0.885$, (e) $E/m=0.875$ and (f) $E/m=0.872$ 
       with $L/m=-7.0M$, $Q_d=70.0M^2$, $r_{\rm ini}=10.5M$ and 
       $\theta_{\rm ini}=\pi/2$.   
   }
 \label{fig:poincare-9} 
\end{figure}

\begin{figure}
  \epsscale{0.8} \plotone{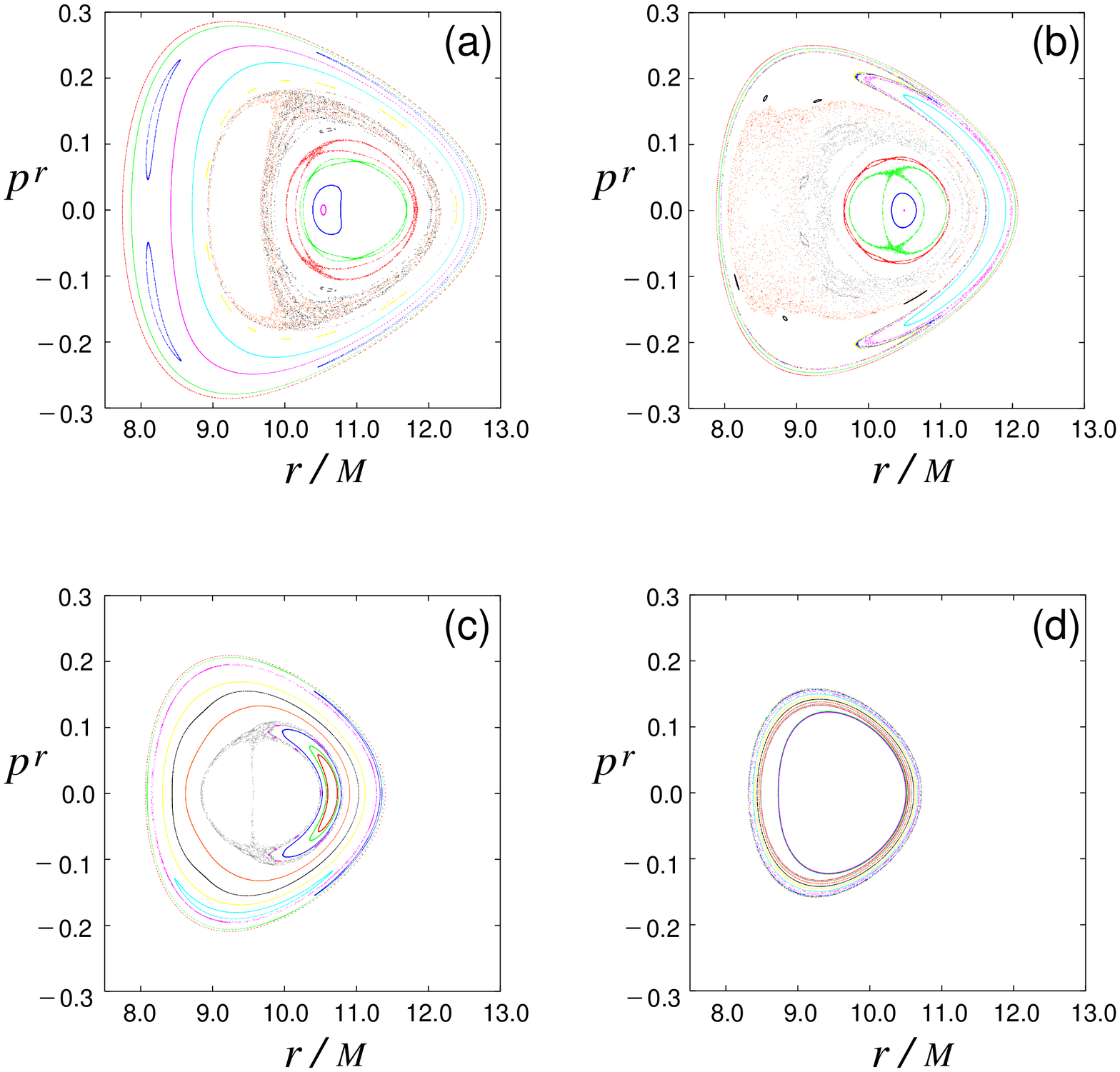}
  \caption{
       The Poincar\'{e} map of a charged particle orbiting in the 
       dipole magnetic field around a black hole of $a=M$. 
       The parameters for the motion are 
       (a) $E/m=0.900$, (b) $E/m=0.890$, (c) $E/m=0.880$ and 
       (d) $E/m=0.870$ with $L/m=-7.0M$, $Q_d=70.0M^2$, 
       $r_{\rm ini}=10.5M$ and $\theta_{\rm ini}=\pi/2$.   
   }
 \label{fig:poincare-M} 
\end{figure}

 Figures~\ref{fig:poincare-3}--\ref{fig:poincare-M} show several
 typical examples of the Poincar\'{e} map for various hole's spin and
 particle's energies.  We find the spin effects on
 the trajectories in the Poincar\'{e} map.  For slowly rotating black
 hole cases, the region of random trajectories fill a finite portion of
 the energy surface in phase space; see Fig.~\ref{fig:poincare-3}.  The
 intersections of a single random trajectory with the surface of section
 fill a finite area.     
 For a rotating black hole cases, both types of trajectories are mixed; 
 see Figs.~\ref{fig:poincare-6} and~\ref{fig:poincare-9}.  That is,
 the distribution on the Poincar\'{e} map makes wide rings by random
 trajectories, and closed curves by regular trajectories.  These
 differences on their trajectories depend on the initial ejection angle
 $\psi_{\rm ini}$.  Specially, for rapidly rotating black hole cases,
 the ratio of the regular orbits increases rather than mildly rotating
 black hole case; see Fig.~\ref{fig:poincare-9}.  The hole's spin
 effects weaken the chaotic motion in the black hole magnetosphere.  In
 fact, for the maximally rotating black hole case, for a wide range of
 $\psi_{\rm ini}$-values the regular orbits are observed, although
 random trajectories also appear for larger energy particle motions; see 
 Fig.~\ref{fig:poincare-M}.

 Next, let us see the energy dependences on the Poincar\'{e} map. 
 For a slowly rotating black hole case (see Fig.~\ref{fig:poincare-3}),
 the random nature of intersections on the Poincar\'{e} map is almost
 independent of particle's energy.  The outer boundary of cross sections
 plotted by the trajectory of $p_\theta^{\rm ini}=0$ only shrinks with
 decreasing the particle's energy; it depends on the effective potential
 well specified by the values of $Q_d$, $L$ and $a$ \citep{KV92}.  We
 also see a few void regions on the map in some cases.     
 On the other hand, we find the energy dependence of the Poincar\'{e}
 map for mildly and rapidly rotating black hole cases.  
 In Figures~\ref{fig:poincare-6}a and~\ref{fig:poincare-6}c (where 
 $a=0.6M$), most trajectories are chaotic, but in
 Figures~\ref{fig:poincare-6}b and~\ref{fig:poincare-6}d the regular  
 trajectories appear on the map (less chaotic). 
 Figure~\ref{fig:poincare-9} and~\ref{fig:poincare-M} show the cases of
 rapidly rotating black hole case.  The regular trajectory is obtained when
 the elevation of the particle ejected is small (somewhat parallel to
 the equator) or large (somewhat perpendicular to the equator), while 
 the random trajectory is observed when the elevation has an intermediate
 value of them.  For example, in Figure~\ref{fig:poincare-9}a, when the 
 initial value of $p^r$ is in the range of A--B or C--D, the regular
 trajectories are obtained, while the range of B--C shows random
 trajectories.

 Although the charged particle is trapped in the effective potential
 well, when the energy has an almost minimum value at the injection
 point; that is, $E/m \sim V_{\rm eff}(r_{\rm ini},\theta_{\rm ini})$,
 the intersections of the whole trajectories are bounded by outer and
 inner closed curves, which make a narrow layer; see
 Figs.~\ref{fig:poincare-9}f and~\ref{fig:poincare-9}d.  
 The inner boundary of the cross section is plotted for the orbit of  
 $p^r_{\rm ini}=0$.  
 Note that, in the case of $a \gtrsim 0.6M$, some orbits with their
 almost minimum energy show the random trajectories on the Poincar\'{e}
 map, but the region of the random trajectories is restricted within a 
 narrow torus-like belt.  The width of this belt becomes narrow when
 the particle's energy has its minimum energy that is specified as the 
 value of the effective potential at the particle's injected point; for
 a slowly rotating black hole case, such a belt-like distribution does
 not appear on the map. 
 In summary, we see the tendency that random trajectories are independent
 of the initial injection angles and their energy in a slowly rotating
 black hole spacetime.  For a rapidly rotating black hole case, however, 
 we also see regular trajectories, which depend on the initial injected
 angle and also their energy.

\section{Discussion}   \label{sec:discuss}

 In \S~\ref{sec:basic-eqs}, we have mentioned that in a stationary and 
 axisymmetric black hole magnetosphere there are only three constants of
 motion, $\pi_t=E$, $\pi_\phi=-L$ and $m$, and as a consequence of the
 non-existence of the forth integral of motion, $\pi_\theta$, the chaos
 would appear in the system.  In spite of being a non-integrable system,
 however, we have found the regular trajectories on the Poincar\'{e}
 map, although we have also seen the chaotic trajectories (for the
 different injection angles).  In this section, we discuss the reason.

 As shown in Figures~\ref{fig:traj_a3}--\ref{fig:traj_a9}, a particle
 in the black hole dipole magnetic field is trapped when a black hole
 rotates.  It bounces back and forth between the northern and southern
 mirrors (also called magnetic bottle). 
 Since the regular or random trajectories depend discontinuously on a 
 choice of initial angle, their presence does not imply the existence
 of a global invariant of the system.  However, if regular trajectories
 exist, they should represent some kind of invariants of the motion.
 These trajectories are conditionally periodic with angle variables.  
 With this motion an adiabatic invariant would be associated, and that
 is related to the longitudinal motion.  Then, we define the action 
 integral as \citep[see, e.g.,][]{GP04} 
\begin{equation}
   J_\theta   
   \equiv \frac{1}{\ell} 
   \oint \ \sqrt{- p_\theta p^\theta } \ d\lambda \ ,
\end{equation}
 where the integral is taken over one cycle of the oscillation in time 
 and $\ell$ is the length of one cycle of the path. 
 In general, the value of $J_\theta$ is not a constant for each cycle of
 the oscillation of a test-charge in the black hole dipole magnetic
 field around a central object, so that we will see random trajectories 
 (chaos).  In fact, for a slowly rotating black hole, we see that most 
 orbits with the various injection angles are chaotic, and their
 $J_\theta$-values are not constants; they vary irregularly.  
 However, we can also observe the regular trajectories, when the action 
 integral of $J_\theta$ is nearly constant for a long-time orbital
 motion.    
 For a rotating black hole case, we can find the regular trajectory
 where the value of $J_\theta$ becomes approximately constant, while 
 it oscillates slightly in a few cycles.  Note that the ``Carter's
 constant ${\cal Q}$'' fluctuates in time; that is, the value of 
 ${\cal Q}$ defined by equation~(\ref{eq:Carter}) is not a constant.   
 In this case, the motion shows a regular trajectory on the
 Poincar\'{e} map.  
 Thus, we find that the motion of a test charge in the magnetosphere
 around a rotating black hole can be nearly integrable.

 \begin{figure}
  \epsscale{1.0}
  \plotone{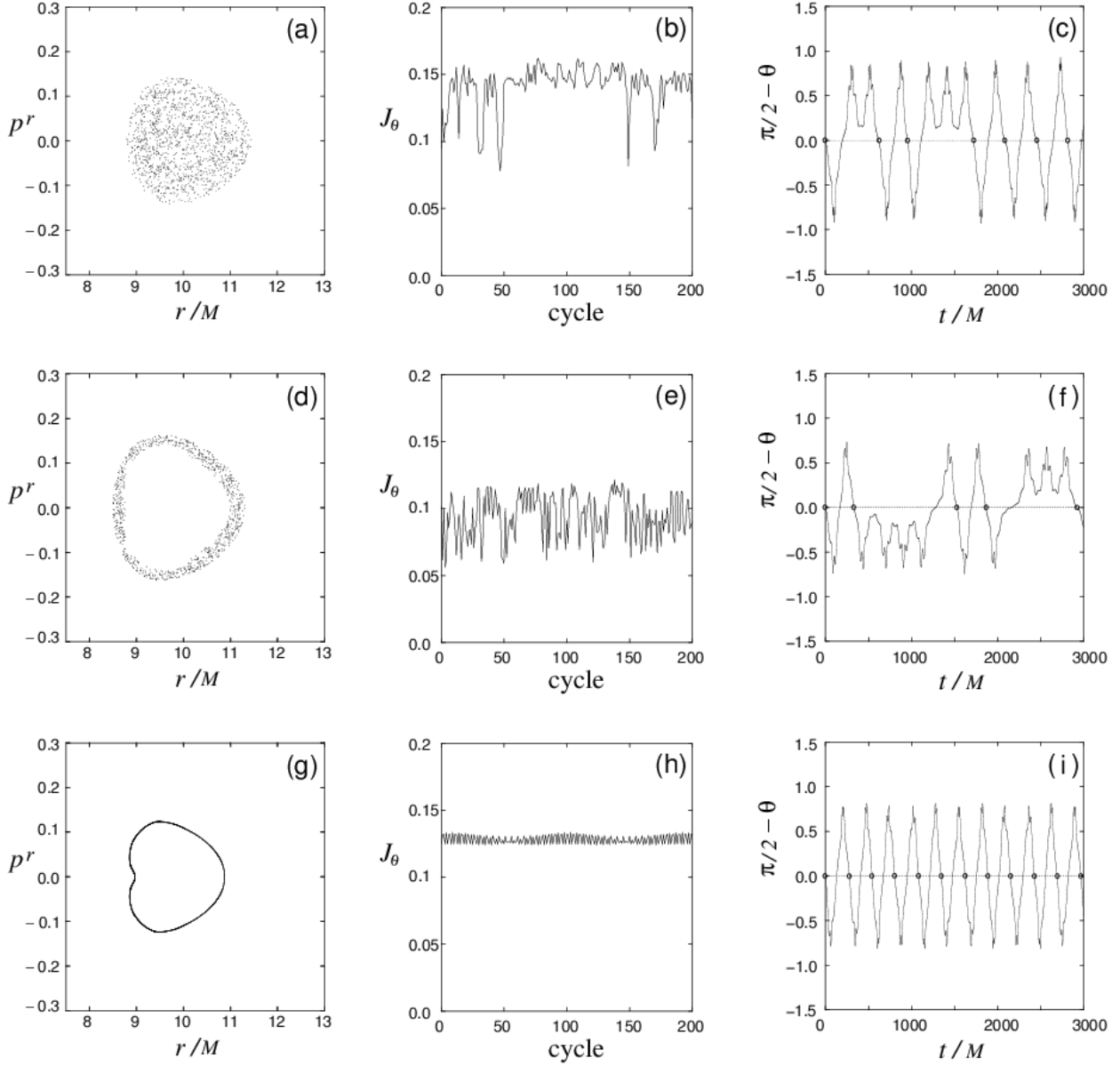}
  \caption{ 
   Various types of the Poincare map (LEFT column).  Many points on
   the Poincare maps (a, d) are plotted by stochasticity (chaos), 
   while that on the Poincare map (g) shows regular trajectory.  
   The action integral $J_\theta$ (CENTER column) and the time series
   of $\pi/2 - \theta$ (RIGHT column), where the first several 
   cycles are shown, are also presented.  Small circles in the time
   series correspond to the points plotted on the Poincar\'{e} map.  
   The values of $J_\theta$ are almost constant for (h) regular
   trajectories.  The parameters of these trajectories are set as  
   (a, b, c) $a=0.3M$, $E/m=0.900$, $\psi_{\rm ini}=-0.10\pi$,     
   (d, e, f) $a=0.6M$, $E/m=0.890$, $\psi_{\rm ini}=-0.10\pi$,     
   (g, h, i) $a=0.6M$, $E/m=0.890$, $\psi_{\rm ini}=-0.30\pi$.  
   The remaining parameters are $L/m=-7.0M$, $Q_d=70.0M^2$, 
   $r_{\rm ini}=10.5M$ and $\theta_{\rm ini}=\pi/2$.    
  }
 \label{fig:type} 
\end{figure} 

 \begin{figure}[t]
 \epsscale{1.0}
 \plotone{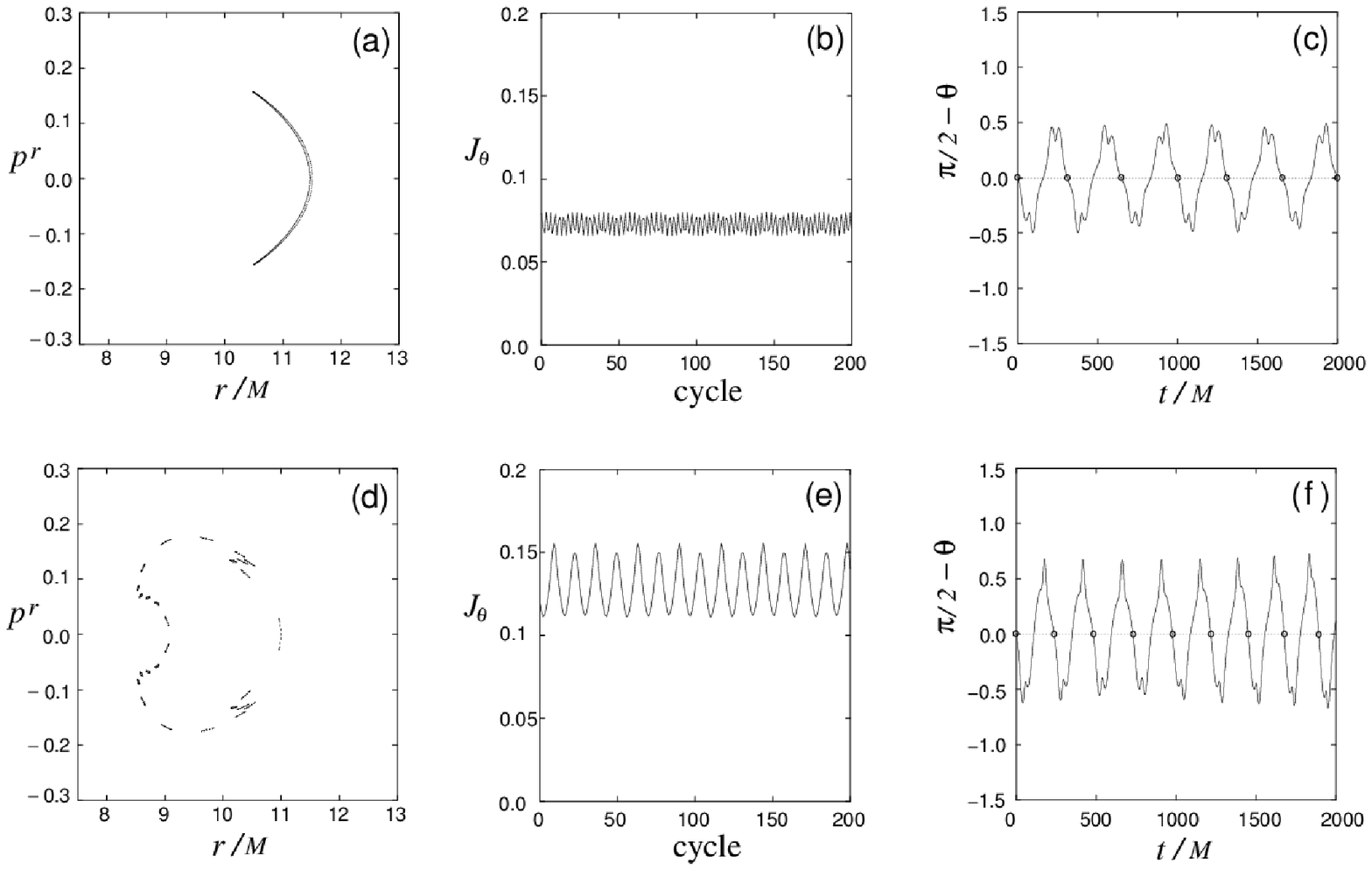}
  \caption{ 
   Examples of the Poincare map (LEFT column) for $a=0.9M$ case. 
   The regular trajectories are shown.  
   The action integral $J_\theta$ (CENTER column) and the time series of 
   $\pi/2 - \theta$ (RIGHT column), where the first several cycles are
   shown, are also presented.  
   The parameters of these trajectories are set as  
   (a, b, c) $\psi_{\rm ini}=-0.10\pi$ and (d, e, f) $\psi_{\rm
   ini}=-0.22\pi$.       
   The other parameters are $E/m=0.885$, $L/m=-7.0M$, $Q_d=70.0M^2$, 
   $r_{\rm ini}=10.5M$ and $\theta_{\rm ini}=\pi/2$.    
  }
 \label{fig:type-a9} 
\end{figure} 

 The region of random trajectories fill a finite portion of the
 surface of a section in phase space.  Figure~\ref{fig:type}a shows that
 the typical intersections of a single random trajectory with the
 surface of a section fill a finite area, where the value of $J_\theta$
 varies irregularly; see Fig.~\ref{fig:type}b.  The time series of
 $\pi/2 -\theta$ also shows an irregular feature in
 Fig.~\ref{fig:type}c. 
 In Figure~\ref{fig:type}d, we see an annular layer randomly filled with   
 intersections of a single trajectory lying between two invariant
 curves.  We also see periodic trajectories that are composed by small
 regular trajectories inside these randomly filled regions (see, e.g.,
 Figs.~\ref{fig:poincare-6}d and~\ref{fig:poincare-9}d).  
 Such a structure may be related to resonances of the orbits that play a 
 crucial role in the appearance of random motions in near-integrable
 systems \citep{LL92}.  In this paper, however, we do not discuss about
 the details.     
 In this annular layer case, we also see that the value of $J_\theta$
 varies irregularly; see Fig.~\ref{fig:type}e. 
 On the other hand, in Figure~\ref{fig:type}g, we see a generic
 trajectory that covers the surface of the torus.  The motion along the
 dipole magnetic field line is almost periodic in the
 $\theta$-direction; see Fig.~\ref{fig:type}i.    
 The intersections of the trajectory with the surface of a section at
 a value of $\psi_{\rm inj}$ lie on a closed invariant curve and densely 
 cover the curve over long periods of time.  The value of $J_\theta$ is
 nearly constant as shown in Figure~\ref{fig:type}h.   
 Figure~\ref{fig:type-a9} shows the other characteristic trajectories in
 a rotating black hole spacetime; see also Fig.~\ref{fig:traj_a9}.  In  
 Figure~\ref{fig:type-a9}(TOP), the crescent regular trajectory is
 presented, where the value of $J_\theta$ is approximately constant. 
 On the other hand, in Figure~\ref{fig:type-a9}(BOTTOM), many islands
 of a trajectory distribute on a curve.  This shows a regular trajectory.
 However, the value of $J_\theta$ is not a constant, and it oscillates
 periodically.  When we consider the average of $J_\theta$ during dozens
 of cycles of oscillations, the value is almost constant over long
 periods of cycles.  It seems that this feature is enough to guarantee
 the generation of the regular trajectory.

 The ratio of the regular trajectories on the Poincar\'{e} map increases
 with the value of the hole's spin.  Specially, for the maximally
 rotating case of $a=M$, almost trajectories become regular; see
 Fig.~\ref{fig:poincare-M}.  It seems that the fourth constant of motion
 is generated for the system with the black hole dipole magnetic field
 in the $a\to M$ limit.  One may expect that, by using the dipole
 magnetic fields (\ref{eq:At_M}) and (\ref{eq:Af_M}), the equation of
 motion can be separable with respect to the $r$ and $\theta$
 coordinates, and the ``modified Carter constant'' including the
 magnetic field terms may be redefined.  However, without the separable
 treatment on the basic equation, we find the constancy of $J_\theta$ 
 in this system numerically.  Exactly speaking, for the regular
 trajectories appeared in this system the value of $J_\theta$ is not
 necessary to be a constant; in fact it oscillates periodically.
 However, for a regular trajectory, we can regard the value of
 $J_\theta$ (or the average on a few cycles) as a constant
 approximately, comparing with that of chaotic trajectories.  Even in
 the $a\to M$ limit, this constancy can be broken depending on the
 injection angle.  In fact, for larger $E/m$, we can also see an annular
 layer of randomly filled by trajectories.

 Our result suggests that the saddle in the double-well potential plays
 an important role for chaotic motions of a charged particle.
 Similar behavior has been investigated in Hamiltonian dynamical systems
 \citep{RZ84}.  
 In the limit of $a\to M$, however, we see that the feature of the
 double well potential vanishes.  Then, the chaotic behavior weakens,
 where many regular trajectories are observed. 
 The property of chaos and/or regular trajectories in the dipole
 magnetic field around a rapidly rotating black hole is related to the
 hole's spin dependence on the effective potential, which is generated
 by the combination of the electromagnetic force, the gravitational
 force and the centrifugal force.  Although in this paper we consider
 the black hole dipole magnetic field, we could understand the basic
 properties of a charged particle by considering the distribution of the
 effective potential when we treat a motion of a test charge in any
 arbitrary electromagnetic field around a black hole.     
 It is a future work to confirm the above suggestion by more detailed
 analysis.

\section{Concluding Remarks}

 In this paper, we have discussed the off-equatorial motion of a test
 charge in the black hole dipole magnetosphere.  The charged particle
 gyrates around a magnetic field line, drifts in the toroidal direction
 and oscillates between the northern and southern hemisphere, and then 
 the orbits becomes very complicated.  So, the numerical study by the
 Poincar\'{e} map is effective.  Then, we find the spin dependence 
 of the trajectory on the Poincar\'{e} map.  That is, for a slowly
 rotating black hole case, the particle's trajectory shows chaos.  
 More interestingly, we also find that, for a rapidly rotating black
 hole case, the chaotic behavior of the trajectories weakens and the
 fourth invariance of the motion, which is an adiabatic invariant
 related to the longitudinal motion, can be approximately generated.

 The chaotic motion in a black hole magnetosphere may be related to the
 origin of cosmic rays.  The magnetic mirror concepts play a prominent
 role in space and plasma astrophysics.  Examples are the Van Allen belt
 in the magnetosphere of the Earth, where high energy particles are
 trapped in a magnetic bottle and some kind of acceleration mechanisms
 are expected.  
 In the case of a black hole magnetosphere, we may expect the formation
 of the magnetic bottle near the black hole.  When such a situation is
 realized, the particles trapped in the ``black hole Van Allen belt''
 would emit high-energy radiation, or some particles may accelerate to
 ultra-relativistic velocity by some electromagnetic interactions (ex.,
 the Fermi acceleration by the perturbed magnetic bottle). 
 Furthermore, some kinds of waves (i.e., Alfv\'en waves and/or fast
 waves) in the magnetosphere will give influence to the motion of a
 particle moving in a magnetized plasma.  
 Our simple model with the black hole dipole magnetic field presented
 here could be a preliminary step toward the understanding the motion of
 charged particles around a black hole.

 In this paper, we do not consider the interactions between charged
 particles, and the emission process from the chaotic/regular motion of
 a charged particle.  However, we can expect that the spectrum emitted
 from such charged particles in periodic motions in the inhomogeneous
 magnetic field carries the informations on the black hole spin and the 
 strength and/or distribution of the electromagnetic field.  
 These studies should be considered in future works.

\acknowledgments

 M.T. would like to thank Sachiko Tsuruta and Rohta Takahashi 
 for their helpful comments.  
 This work was supported in part by the Grants-in-Aid of the Ministry 
 of Education, Culture, Sports, Science and Technology of Japan
 (19540282, M.T.).




Facilities: ---

\end{document}